# From Playability to a Hierarchical Game Usability Model

Lennart Nacke
School of Computing
Blekinge Institute of Technology, Sweden
Lennart.Nacke@acm.org

## ABSTRACT
This paper presents a brief review of current game usability models. This leads to the conception of a high-level game development-centered usability model that integrates current usability approaches in game industry and game research.

## Categories and Subject Descriptors
K.8.0 [**General**]: Games – *Personal Computing.* H.5.1 [**Multimedia Information Systems**]: Evaluation/methodology - *Information Interfaces and Presentation.* J.4 [**Computer Applications**]: Psychology.

## 1. INTRODUCTION
In recent years, we are seeing a lively debate about measures for usability and user experience in game contexts [1]. Admittedly, usability and playability may be dichotomous concepts [5, 7], but to understand how they relate to each other, we need to establish a basic taxonomical understanding of each term first.

Usability research generally operates within the ISO 9241-11 usability standard, which defines usability as the *"extent to which a product can be used by specified users to achieve specified goals with effectiveness, efficiency, and satisfaction in a specified context of use"* [4]. This conception of usability defines measurable factors. This led Sauro [11] to conclude that *effectiveness* can be measured in completion rates and errors, *efficiency* from time on task, and *satisfaction* using any of a number of standardized satisfaction questionnaires. This puts a numerical fundament under usability studies. Following this argumentation, one could see digital games merely as software with the same requirements on interaction as other products. However, experiencing gameplay does not equal experiencing regular product use. Thus, assuming that playability is related to measuring an experience, Fabricatore et al. find that *"playability is the instantiation of the general concept of usability […] determined by […] understanding and controlling gameplay"* [3]. This could be interpreted to support a dichotomy of playability and usability, saying that playability is related only to a game's *"capability to provide enjoyment for a player over an extended period of time"* [7]. Since measurable factors of gameplay are being discovered for objective quantitative measurement [8, 9, 13], these need to be incorporated in future playability definitions.

## 2. MODELS OF PLAYABILITY
Järvinen et al. [5] establish a definition of playability for both design and evaluation: *"a collection of criteria with which to evaluate a product's gameplay or interaction"*. They proceed to introduce four components of playability derived from their take on flow in games [12]: (1) functional, (2) structural, (3) audiovisual, and (4) social playability, which are then analyzed in regard to *formal* aspects (game functional, e.g. rules) and *informal* aspects (user experience).

Functional playability maps input peripherals to requirements of gameplay, essentially referring to the learning curve of a game's input/output modalities. This relates to efficiency in the standard usability definition [4], but refers to interaction functionalities in game context. Structural playability is an expert evaluation of game rules, structures and patterns as well as player evaluation of skill, experience, and actions. This can be compared to usability heuristics evaluation of products [10]. Audiovisual playability refers to the quality of the visualization and sonification of game elements. It is naturally tied to functional playability as interface aspects can directly relate to input controls and feedback of the game. Social playability estimates the suitability of digital games for different contexts of use, using a lens through which game cultures are examined that foster a sense of community. This aspect of playability can only be assessed using long-term anthropological or sociological studies. While those four components of playability present a general approach to evaluating game experience, only functional and audiovisual playability are concepts that can be quantified using metric evaluation methods. What this approach [5] lacks are concrete examples of empirically measurable criteria of playability.

Desurvire et al. [2] present four game heuristic categories: *game play*, *game story*, *game mechanics*, *game usability*. The first category (game play) is in line with the descriptions of Korhonen's *gameplay* [6] and Järvinen's *structural playability* [5]. Game story certainly looks at a game's narrative development (if existent). Game mechanics are interface and programming rules, together with game usability, they relate to Korhonen's notion of game usability [6] and Järvinen's functional and audiovisual playability [5]. Korhonen et al. [6] presented playability heuristics modules for *game usability*, *mobility* and *gameplay*. Their heuristics for game usability are in line with what Järvinen et al. [5] call functional and audiovisual playability evaluation. The gameplay heuristics relate to the structural content of a game, placing them in a similar category as structural playability [5].

## 3. A GAME USABILITY MODEL
Most models acknowledge that usability is a fundament of playability, which leads us to propose the following model.

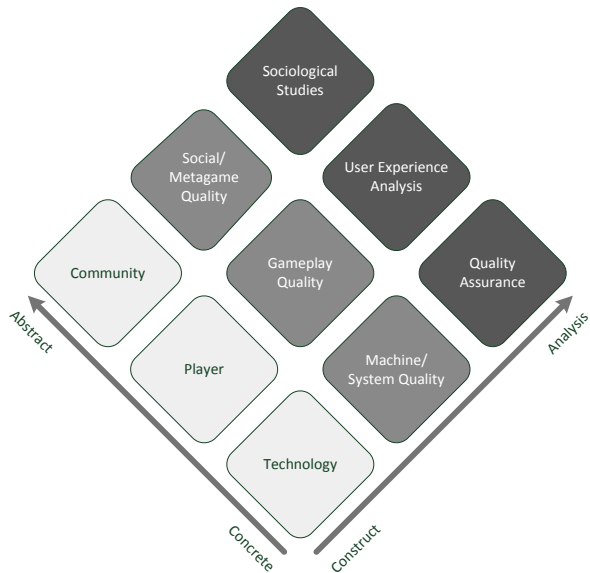

**Figure 1. A hierarchical model of game usability**

Figure 1 shows a hierarchical game usability model that accounts for *concrete* to *abstract*, *measurable* entities, which are described from *theoretical* construct to *practical* application. Working technology is the foundation of any digital game and it refers to the quality of the system. Quality assurance for it can be done using industry standard playtesting and bug tracking. In any way it refers to numerical evaluations of technological functionality, incorporating functional playability and game usability notions [2, 5, 6]. The level of abstraction is player-focused evaluation, which assesses the gameplay experience using subjective and objective measures [8, 9]. It can be placed in the context of gameplay and structural playability evaluation [2, 5, 6]. Finally, we have to take into account reception and cultural significance of a game, which can also be measured abstractly using anthropological and sociological approaches. This correlates with social playability and is something that is usually rather done in research than in industry context. We will continue to expand this model in future research, aiming to exemplify its applicability to all phases of game development.

## 4. ACKNOWLEDGMENTS
This research was funded by the European Commission under the 6th Framework Programme: NEST: FUGA (Contract: FP6-NEST-28765).